\documentclass[letterpaper, onecolumn]{autart}  % Comment this line out
\usepackage{epsfig} % for postscript graphics files
\usepackage{subfigure}
\usepackage{amsmath} % assumes amsmath package installed
\usepackage{amssymb}  % assumes amsmath package installed

\usepackage{amsfonts}
\usepackage{enumerate}
\usepackage{color}

\newcommand{\field}[1]{\mathbb{#1}}

\newcommand{\R}{\field{R}}

\newcommand{\co}{\mathrm{conv}}
\newcommand{\trace}{\mathrm{trace}}

\newtheorem{lemma}{Lemma}
\newtheorem{thrm}{Theorem}
\newtheorem{problem}{Problem}

\newtheorem{remark}{Remark}

\begin{document}

\begin{frontmatter}

%\title{\LARGE\bf Guaranteed Cost Control for Uncertain  Quadratic Systems}
\title{\LARGE\bf Optimal Control of Uncertain Nonlinear Quadratic Systems with Constrained Inputs}

\thanks[footnoteinfo]{This paper was not presented at any IFAC
meeting.}

\author[UMG]{Alessio~Merola}\ead{merola@unicz.it},
\author[UMG]{Carlo~Cosentino\corauthref{cor}}\corauth[cor]{Corresponding author, Tel.:
+39 09613694051, Fax: +39 09613694090.}\ead{carlo.cosentino@unicz.it},
\author[UMG]{Domenico~Colacino}\ead{colacino@unicz.it},
\author[UMG]{Francesco~Amato}\ead{amato@unicz.it}

\address[UMG]{School of Computer and Biomedical Engineering, Universit\`a degli Studi Magna Gr{\ae}cia di Catanzaro, Campus Universitario di Germaneto, 88100
Catanzaro, Italy}

                             % keyword list or with the
                                          % help of the Automatica
                                          % keyword wizard
\begin{abstract}
This paper addresses the problem of robust and optimal control
for the class of nonlinear quadratic systems subject to norm-bounded parametric uncertainties and disturbances, and in presence of some amplitude constraints on the control input. By using an approach based on the guaranteed cost control theory,  a technique is proposed to design a state feedback controller ensuring for the closed-loop system: i) the local exponential stability of the zero equilibrium point; ii) the inclusion of a given region into the domain of exponential stability of the equilibrium point; iii) the satisfaction of a guaranteed level of performance, in terms of boundedness of some optimality indexes.  In particular, a sufficient condition for the existence of a state feedback controller satisfying  a prescribed integral-quadratic index is provided, followed by a sufficient condition for the existence of  a state feedback controller satisfying a given $\mathcal L_2$-gain disturbance rejection constraint. By the proposed design procedures, the optimal control problems dealt with here can be efficiently solved as Linear
Matrix Inequality (LMI) optimization problems.
%Furthermore, a nontrivial application of the proposed approaches, namely the regulation and stiffness control of a biomimetic joint, is illustrated;
%such application shows the potential benefits of the design procedures, among others,  for mechatronic and robotic applications.
\end{abstract}

\end{frontmatter}

%%%%%%%%%%%%%%%%%%%%%%%%%%%%%%%%%%%%%%%%
\section{Introduction} \label{sec:intro}
%%%%%%%%%%%%%%%%%%%%%%%%%%%%%%%%%%%%%%%%
The goal of this paper is to investigate the extension of the linear quadratic regulator (LQR) and $\mathcal H_\infty$ optimal control techniques to the class of nonlinear quadratic systems (NQSs).\\
%
%NQSs play an important role to explain the dynamic behavior of several phenomena both in engineering (electric power systems~\cite{Jyotika}, chemical reactors~\cite{Rossler}, and robots~\cite{Sciavicco}), as well as in other areas such as biology, ecology, economics and meteorology~\cite{Lorenz:63}. In biology, some quadratic laws describe the interaction dynamics between different (biochemical and biological) species, e.g., in enzymatic reaction kinetics \cite{Murray}, or in Lotka--Volterra prey predator models \cite{Lotka,Volterra}.
%
The stability analysis and design of nonlinear quadratic systems has been performed in \cite{Amato:Automatica,Amato:CDC07,Amato:TAC:2010}; these papers provide conditions ensuring the existence of state feedback controllers, which stabilize the given quadratic system and guarantee that an assigned polytopic region belongs to the domain of attraction of the zero equilibrium point; applications of such approach are reported in \cite{Merola:BSPC}, to study the interaction dynamics between tumor and immune system, and in \cite{JCB}, to investigate the bistable behavior of gene regulatory network.

The extension of the above-mentioned optimal control methodologies to NQSs will be pursued through an approach that is reminiscent of the  Guaranteed Cost Control (GCC) theory \cite{Chang:72}.%, which is exploited to solve the LQ control problem in presence of plant uncertainties.
%In the seminal work~\cite{Petersen:94}, a Riccati equation approach to the design of optimal quadratic guaranteed cost controllers for uncertain linear systems is proposed.
GCC-based methodologies guarantee that the control performance is bounded by a specified performance level for all admissible uncertainties of the closed loop system~\cite{Petersen:94,Costa:02}.\\%,Zhang:09}.
%
%However, LQ control is inadequate  arising, e.g., both from the incomplete knowledge of the plant model and the parameter variability. Robust control can be efficiently %addressed in the framework of optimal control theory through the.
%When dealing with parameter uncertainty, the GCC approach can provide significant results which overcome the limitations of LQ optimal control.
%
%Therefore, in light of the potential applications, the robust and optimal control for  uncertain systems may be of great interest for the control engineering community.
%
In the GCC literature, few works have dealt with nonlinear systems; for instance, in~\cite{Aliyu:2000}, a minimax optimization methodology has been developed for designing a robust GCC  law for a class of uncertain nonlinear systems, whereas some LMI-based conditions have been formulated in~\cite{Coutinho:02} to solve a robust GCC problem for a class of input-affine nonlinear systems.  Preliminary works concerning guaranteed-cost optimal control of NQSs can be found in~\cite{Amato_RAS:12,Amato:ECC14}.\\
As $\mathcal H_\infty$  optimal control theory for nonlinear systems is concerned, the design of state feedback controllers is tackled in~\cite{Kim:2005}, where
bilinear systems are considered, whereas~$\mathcal H_\infty$ filtering for a class of Lipschitz nonlinear systems with time-varying
uncertainties is proposed in \cite{Abbaszadeh:2012}, in order to attain both the exponential stability
of the estimation error dynamics and robustness against uncertainties.
In~\cite{Jianbin:2011}, the $\mathcal H_\infty$ control theory has been extended to the class of discrete-time
piecewise-affine systems with norm-bounded uncertainties; the basic aim of the contribution is to design
a piecewise-linear static output feedback controller guaranteeing the asymptotic
stability of the resulting closed-loop system with a prescribed
$\mathcal H_\infty$ disturbance attenuation level.

Since the achievement of global stabilization and/or the determination of the optimal cost  is a difficult or even impossible task when NQSs are dealt with, following the guidelines of  \cite{Amato:Automatica,Amato:CDC07,Amato:TAC:2010},  we look for sub-optimal controllers with guaranteed performance into a certain compact region containing the origin of
the state space (such region  can be interpreted as the operating domain of the system). More precisely,  given an uncertain NQS, possibly subject to exogenous disturbances, the main results of this paper consist of some sufficient conditions for the existence of a linear state feedback controller which will ensure for the closed-loop system: i) the local exponential stability of the zero equilibrium point; ii) the inclusion of a given  region into the domain of exponential stability of the equilibrium point itself; iii) the safisfaction of a guaranteed level of performance, in terms of the boundedness of a quadratic cost function in the form
\[
\int_0^{\infty} \left(x^TQx +u^TRu\right)dt\,,
\]
where $x$ and $u$ are the system input and state, respectively (when the extension of the LQR approach is considered), or in terms of the negativeness of a quadratic cost function in the form
\[
 \int_0^{\infty} \left(z^Tz -  w^Tw\right)dt\,,
\]
where $z$ and $w$ are the system controlled variable and the disturbance, respectively (when the $\mathcal H_\infty$ case is considered).

%Then, the $\mathcal H_\infty$ control problem is considered; in this case, the proposed
%methodology allows to design a linear state feedback control law, which guarantees a desired $\mathcal L_2$-gain disturbance rejection performance.

It is worth noting that the proposed results, for both  optimal control problems, can explicitly take into account assigned constraints on the control input amplitude.

The devised conditions involve the solution of  LMI optimization problems, which can be efficiently solved via off-the-shelf routines.
%A nontrivial application, involving the experimental modeling, identification and control of a robot joint actuated by pneumatic artificial muscles, is illustrated to show the effectivenes of the proposed techniques.

The remainder of the paper is organized as follows. Section~\ref{sec:prob_stat} provides the problems statement and some preliminary results.
The main results of the paper, namely some sufficient conditions for the existence of linear state feedback controllers guaranteeing optimal quadratic regulator and~$\mathcal H_\infty$ performance, are presented in Section~\ref{sec:main}.
%The proposed approach is then applied in Section~\ref{sec:results} to design the control system for  a biomimetic robot joint actuated by pneumatic artificial muscles, whose model is derived in Section~\ref{sec:model}.
Eventually, some concluding remarks are given in Section~\ref{sec:concl}.

\emph{Notation:} The symbol $\mathcal L_2^{n_w}$ denotes the subspace of vector-valued functions in $\mathbb R^{n_w}$ which are square-integrable over $[0,+\infty)$ with Euclidean vector norm $||\cdot||_2=(\int_0^{\infty} ||\cdot||^2 dt)^{1/2}$. The matrix operation $A\otimes B$ denotes the Kronecker product of matrices $A$ and $B$, while $I_n$ denotes the identity matrix of order $n$. Given a square matrix $M$, $\mathrm{symm}(M) := M + M^T$.  In general, when it is not explicitly specified, all matrices must be intended of compatible dimensions.

%%%%%%%%%%%%%%%%%%%%%%%%%%%%%%%%%%%%%%%%%%%%%%%%%%%%%%%%%%%%%%%%%%%%%%%%%%%%%%
\section{Problem statement and preliminaries}\label{sec:prob_stat}
%%%%%%%%%%%%%%%%%%%%%%%%%%%%%%%%%%%%%%%%%%%%%%%%%%%%%%%%%%%%%%%%%%%%%%%%%%%%%%

%%%%%%%%%%%%%%%%%%%%%%%%%%%%
\subsection{Uncertain NQSs}
%%%%%%%%%%%%%%%%%%%%%%%%%%%%
Consider the class of  uncertain NQSs, described by the following state-space representation
\begin{equation}\label{eq:sys}
\begin{split}
\dot x(t)&=(A+\Delta A)x(t)+f(x(t))+\Delta f(x(t))\\
&+(B+\Delta B)u(t)+g(x(t),u(t))\\
&+\Delta g(x(t),u(t))+ B_w w(t)\\
%x(0) &=x_0\,,
 z(t)&=C x(t)\,,
\end{split}
\end{equation}
where $x(t)\in\R^n$ is the system state, $u(t)\in \R^m$ is the control input, $z(t)\in \R^{n_z}$ is the controlled variable,
$w(t)$ denotes the external disturbance which belongs to the space of square-integrable functions $\mathcal L_2^{n_w}[0,+\infty)$.
It is assumed that the energy of the disturbance is bounded, that is $\|w\|_2^2\leq 1$.
%\begin{equation} \label{w_bound}
%\|w\|_2^2\leq 1.
%\end{equation}

The matrices $\Delta A$ and $\Delta B$ describe the uncertainties of the linear part of system \eqref{eq:sys}.
The nonlinear and uncertain dynamics are described by the vector-valued functions
\begin{equation}\label{eq:N_x_u}
\begin{split}
f(x)&=\begin{pmatrix} F_1^T x & F_2^T x & \dots & F_n^T x \end{pmatrix}^T x\,,\\
\Delta f(x)&=\begin{pmatrix} \Delta F_1^T x & \Delta F_2^T x & \dots \Delta F_n^T x \end{pmatrix}^T x\,,\\
g(x,u)&=\begin{pmatrix} G_1^T x & G_2^T x & \dots & G_n^T x \end{pmatrix}^T u\,,\\
\Delta g(x,u)&=\begin{pmatrix} \Delta G_1^T x & \Delta G_2^T x & \dots \Delta G_n^T x \end{pmatrix}^T u\,,
\end{split}
\end{equation}
where $F_i\in \mathbb{R}^{n\times n}$, $G_i\in \mathbb{R}^{n\times m}$ $i=1,\dots,n$, are known constant matrices, whereas $\Delta F_i$, $\Delta G_i$ $i=1,\dots,n$, denote parameter-varying matrices of appropriate dimensions.

%$N_i\in\R^{n\times m}$,

%Without loss of generality, it is assumed that system \eqref{eq:sys} is subject to norm-bounded parametric uncertainties of the form
It is assumed that the uncertainties in \eqref{eq:sys} exhibit a structured, norm bounded form, that is
\begin{align}\label{eq:def_NBUs}
\left[%
\begin{array}{cccccccc}
   \Delta A & \Delta B & \Delta F_1 & \dots & \Delta F_n & \Delta G_1 & \dots & \Delta G_n
\end{array}%
\right]=&\nonumber\\
&\hspace{-7.7cm}D H \left[
\begin{array}{cccccccc}
   E_1 & E_2 & R_1 & \dots & R_n & S_1 & \dots & S_n
\end{array}
\right]\,,
\end{align}
where  $H$ is any matrix\footnote{Without loss of generality, $H$ can be any Lebesgue measurable time-varying matrix-valued function (see \cite{Amato}).}
 satisfying $H^TH\leq I$.
%\begin{equation}\label{eq:def_F}
%H^TH\leq I\,.
%\end{equation}
As usual, $I$ denotes any identity matrix of compatible dimensions and $D$, $E_1$, $E_2$, $R_1, \dots R_n$, $S_1, \dots S_n$ are known constant matrices of appropriate dimensions.
Furthermore, the following set of constraints on the control input of system \eqref{eq:sys} is specified
\begin{equation}\label{eq:cstr_u}
|u_i(t)|\leq u_{i,max}\,,
\end{equation}
where $u_{i,max}$, $i=1,\dots,m$ denote prescribed peak bounds on each component of $u(t)$.
%
%\begin{remark}
%The structured norm-bounded uncertainties are widely used in optimal and robust control of linear systems encompassing a large number of applications.
%For instance, such uncertainties can embody the effects of modelling mismatch, nonlinearities due to friction, force disturbance occurring in servomechanisms.
%\end{remark}
%
%\begin{remark}
%An important issue of all control systems is the limitation of the magnitude of the control inputs due to the range of operation of the actuators.
%Therefore, some amplitude constraints on the control inputs have also to be taken into account in the design of optimal controllers.
%\end{remark}
%
%%%%%%%%%%%%%%%%%%%%%%%%%%%%%
\subsection{Problems statement}
%%%%%%%%%%%%%%%%%%%%%%%
The present work investigates the state feedback control problem for system \eqref{eq:sys}; more precisely, we focus on linear state feedback controllers in the form
\begin{equation}\label{eq:sf}
u(t)=Kx(t)\,,
\end{equation}
where $K \in \mathbb{R}^{m\times n}$ is the control gain matrix.
The reason for considering linear controllers is twofold. First of all, linear design permits a very simple implementation of the control system; moreover, as we shall show later, it allows to derive a convex optimization procedure for the selection of the optimal controller gain matrix.\\
The resulting closed loop system has the following form
\begin{equation}\label{eq:cl_sys}
\begin{split}
\dot x&=\Bigl(A + BK + DH \bigl(E_1 + E_2 K \bigr) \Bigr)x\\
&+\begin{pmatrix} (F_1+D H R_1)^T x &\dots& (F_n+D H R_n)^T x \end{pmatrix}^T x\\
+&\begin{pmatrix} K^T(G_1+D H S_1)^T x &\dots& K^T (G_n+D H S_n)^T x \end{pmatrix}^T x\\
&+ B_w w\,.
\end{split}
\end{equation}
In the following, letting $B_w=0$, if the controller $K$ is such that the closed loop system~\eqref{eq:cl_sys} is (locally) exponentially stable for all admissible uncertainties,  we refer to the  {\em domain of exponential stability} of system~\eqref{eq:cl_sys} (DES)\footnote{For the sake of simplicity, we adopt the statement  {\em the DES of the closed loop system} in place of  {\em the DES of the zero equilibrium point of the closed loop system}.} as the  connected set surrounding the origin, such that any trajectory starting at a point in the DES converges exponentially to zero for all admissible uncertainties.

%%%%%%%%%%%%%%%
\subsubsection{Extension of the LQR methodology to NQSs}
%%%%%%%%%%%%%%%%%

Consider the quadratic cost function
\begin{equation}\label{eq:cost_idx}
J_2:= \int_0^{\infty} \left(x^T(t)Qx(t) +u^T(t)Ru(t)\right)dt\,  \,,
\end{equation}
associated to the closed loop system \eqref{eq:cl_sys} with $B_w\equiv0$, where $Q$ and $R$ are symmetric positive definite matrices. It is well known that,
by a proper choice of the weighting matrices $Q$ and $R$, it is possible to specify the desired
quadratic-regulator control performance.

Note that the cost index function \eqref{eq:cost_idx} depends on  the control input as well as on the initial conditions.
By assigning a closed set $\mathcal D\subset \R^n$, $0\in \mathcal D$, the designer is allowed to specify the operative range
in the state space over which the control performance has to be guaranteed.
In the following we shall refer to $\mathcal D$ as the {\em admissible} set.
%\begin{remark}
%It is worth noting that we talk of the {\em extension} of the $\mathcal H_2$ approach to the class of NQSs, rather than the {\em solution} of the $\mathcal H_2$ problem for NQSs. This is due to the fact that, from one hand, we propose a sub-optimal solution to the control problem, and, from the other hand, it is not correct to identifying the minimization of the cost index $\eqref{eq:cost_idx}$ as the solution of an $\mathcal H_2$-type problem. Indeed $\mathcal H_2$ optimization refers to the frequency domain interpretation of the cost index, which does not make sense in the nonlinear context. Similar considerations can be repeated for the $\mathcal H_\infty$ problem discussed in the next section.
%\end{remark}
In the sequel, the definition of Quadratic Guaranteed Cost Controller (QGCC) for the class of uncertain NQSs is precisely stated.
\begin{defn}\label{defn:RGCC}
Consider the NQS \eqref{eq:sys} %with the admissible uncertainties satisfying \eqref{eq:def_NBUs}-\eqref{eq:def_F}, and
with~${B_w=0}$. Given the cost
function~\eqref{eq:cost_idx}, an admissible set~$\mathcal D$, and a positive definite matrix~$P$, the static state feedback controller
\eqref{eq:sf} is said to be a QGCC, with associated cost matrix $P$, for the uncertain system~\eqref{eq:sys} if the following hold:
\begin{itemize}
\item[\textbf{i)}] The admissible set $\mathcal D$ is included into the DES of the closed loop system
\eqref{eq:cl_sys};
\item[\textbf{ii)}] The  performance index~\eqref{eq:cost_idx} for the closed loop system~\eqref{eq:cl_sys} satisfies, for all $x_0\in\mathcal D$, and for all~$H^T H \leq I$%$H$ satisfying~\eqref{eq:def_F},
\[
J_2\leq x_0^TPx_0\,.
\]
\end{itemize}
\end{defn}%\hfill$\triangle$

\begin{remark}
The term quadratic in Definition \ref{defn:RGCC} follows from the fact that, according to condition ii), we require that the cost is bounded by a quadratic form of the initial state. This is consistent with the GCC theory developed for linear systems (see  \cite{Petersen:94}).
\end{remark}

It is worth noting that condition i) in Definition \ref{defn:RGCC} guarantees that the trajectory of the closed loop system starting at any point $x_0\in \mathcal D$ exponentially converges to zero, which in turn implies well posedness of condition ii).

%%%%%%%%%%%%%%%%%%%%%%%%%%%%%%%%%%%
\subsubsection{Extension of the  $\mathcal H_\infty$ optimal control to NQSs}
%%%%%%%%%%%%%%%%%%%%%%%%%%%%%%%%%

The problem of conferring robustness to the closed loop system subject to disturbance input is considered here. A state feedback controller in the form \eqref{eq:sf},
attenuating the effects of the exogenous disturbance signals on the system response, can be designed resorting to an $\mathcal H_\infty$-like control theory.
In this framework, disturbance attenuation can be achieved through the cost function
\begin{equation}\label{eq:cost_idx_H_inf}
J_\infty:= \int_0^{\infty} \left(z^T(t)z(t) - w^T(t)w(t)\right)dt\,.
\end{equation}
Note that the cost index \eqref{eq:cost_idx_H_inf} depends on the control input $u$, and the exogenous disturbance $w$; according to the $\mathcal H_\infty$ framework, the initial state is assumed to be zero. % and the weighting matrices are embedded into the system matrices.

The extension of the $\mathcal H_\infty$ control problem to NQSs can be easily recast in terms of $\mathcal L_2$-gain \cite{Vanderschaft:1992}; indeed the existence of a state feedback control law in the form \eqref{eq:sf} such that $J_\infty <0$ for all $w \in \mathcal L_2^{n_w}[0,+\infty)$, $w(t):\|w\|_2^2\leq 1$,
%with $w(\cdot)$ satisfying \eqref{w_bound}, and all
 and $H^T H \leq I$
%$H$ satisfying ~\eqref{eq:def_F},
implies that, for all admissible uncertainties,
\begin{equation} \label{L2_interpretation}
\sup_{w \in \mathcal L_2^{n_w}[0,+\infty)\atop{\|w\|_2\leq 1}} \frac{\| z \|_2}{\|w\|_2}  < 1 \,.
\end{equation}

The left hand side in \eqref{L2_interpretation} can be interpreted as the~$\mathcal L_2$-gain of the NQS \eqref{eq:sys}; therefore negativeness of $J_\infty$ implies that the $\mathcal L_2$ gain of the NQS \eqref{eq:sys} is guaranteed to be less than 1 for all admissible uncertainties. This justifies the following definition.

\begin{defn}\label{defn:H_inf}
Consider the NQS \eqref{eq:sys}. %with the admissible uncertainties satisfying \eqref{eq:def_NBUs}-\eqref{eq:def_F}.
Given the cost function~\eqref{eq:cost_idx_H_inf}, the static state feedback controller \eqref{eq:sf} is said to be
a guaranteed $\mathcal L_2$-performance controller (G$\mathcal L_2$PC), for the uncertain quadratic system \eqref{eq:sys} if
\begin{itemize}
\item[\textbf{i)}]  The closed loop system \eqref{eq:cl_sys} is (locally) exponentially stable for any matrix $H$ such that $H^T H \leq I$;
%satisfying \eqref{eq:def_F};
\item[\textbf{ii)}] Starting from zero initial conditions,  the $\mathcal L_2$-performance of the closed loop system \eqref{eq:cl_sys} satisfies, for all matrix $H$ such that $H^T H \leq I$,% $H$ satisfying \eqref{eq:def_F},
\[
\sup_{w \in \mathcal L_2^{n_w}[0,+\infty)\atop{\|w\|_2\leq 1}}J_\infty < 0 \,.
\]
\end{itemize}
\end{defn}%\hfill$\triangle$

%\begin{remark}
%Note that, in Definition \ref{defn:H_inf}, the disturbance $w$ plays a role analougous to that one of the initial state $x_0$ in Definition \ref{eq:cost_idx}. Indeed both $w$ and $x_0$ are constrained to belong to compact sets (this explains to use of the $\max$ operator in the  definition of $J_2$ and $J_\infty$). The assumptions on $x_0$ and $w$ is necessary, in the nonlinear cntext, in order to obtain operative results. In the linear context, there is no need of

In the following we denote by $\mathcal R$ the reachable set associated
to the uncertain NQS \eqref{eq:sys}, that is
\begin{align*}%\label{reachabilityset}
\mathcal R &:= \left\{ x(T)\in\R^n \,:\, x(\cdot),\, w(\cdot) \; \mathrm{satisfy}\; \eqref{eq:sys}, \, T\geq 0 \,,\right.\\
%\nonumber \\
 & \hspace{2cm} \left. x(0)=0\,, H^T H\leq I\,, \|w\|_2^2\leq 1 \right\} \,.
\end{align*}
%is instrumental to the solution of the $\mathcal L_2$-performance control problem.
%\begin{defn}
%A connected set $\mathcal R \subset \mathbb R^n$ is said to be a \emph{reachable set} for system \eqref{eq:sys} if, for any $w \in \mathcal L_2^{n_w}[0,+\infty)$, with $w(\cdot)$ satisfying \eqref{w_bound},
%\[
%x(0)=0 \implies x(t)\in \mathcal R\,,\forall t\geq 0\,.
%\]
%\end{defn}%\hfill$\triangle$
According to the above definition, the set $\mathcal R$ envelopes all the trajectories which, starting from zero initial conditions, are perturbed by an admissible exogenous bounded-energy disturbance signal. %In the following, we assume that $\mathcal R$ is a finite subset of $\R^n$.
\begin{remark}
Condition {\bf i)} in Definition \ref{defn:H_inf} plays a role analogous to the internal stability requirement in the context of the $\mathcal H_\infty$ control of linear systems. To this regard, note that condition {\bf ii)} alone does not guarantee  exponential stability of the closed loop system, since some unstable open loop dynamics might be not included in the index \eqref{eq:cost_idx_H_inf}. Also, condition {\bf i)} guarantees that the DES of the closed loop system does not reduce to a singleton; later in the paper we shall see that the reachability set of the closed loop system (if finite) is an estimate of the DES.
\end{remark}
%
%\vspace{-0.3 cm}
%

%%%%%%%%%%%%%%%%%%%%%%%%%%%%%
\subsection{Some ancillary results}
%%%%%%%%%%%%%%%%%%%%%%%%%%%%%%
%
%\vspace{-0.3 cm}
%
Before introducing the main results on the design of QGCCs and G$\mathcal L_2$PCs for the  uncertain NQS~\eqref{eq:sys} with input constraints~\eqref{eq:cstr_u}, some preparatory results are necessary.
First, we recall the following lemma, whose proof can be easily derived from the result in \cite{Petersen:87}.
%
%\vspace{-0.4 cm}
%
\begin{lemma}\label{lemma:petersen}
Given any scalar $\epsilon >0$, some matrices of appropriate dimensions $\Omega_1$, $\Omega_2$, $\Omega_3$
and any matrix $\mathcal M$ such that $\mathcal M^T \mathcal M \leq I$, then
\begin{align*}
%x^T \Omega_1 \mathcal M \Omega_2 x + x^T \Omega_2^T \mathcal M^T \Omega_1^T x & \leq \\
\mathrm{symm} \bigl(x^T \Omega_1 \mathcal M \Omega_2 x \bigr) & \leq\\
\epsilon x^T \Omega_1 \Omega_1^T x &+\epsilon^{-1} x^T \Omega_2^T \Omega_2 x\,,\forall x \in \R^n\,.\\
\end{align*}
\end{lemma}
%
%\vspace{-1 cm}
%
%ERROR HERE
\begin{lemma}\label{lemma:bound}
Consider the uncertain NQS \eqref{eq:sys}, with $B_w=0$.%, and the admissible uncertainties satisfying \eqref{eq:def_NBUs}-\eqref{eq:def_F}.
Given an admissible set~$\mathcal D$ and the cost index~\eqref{eq:cost_idx},
assume there exist some positive scalars $\epsilon_1$, $\epsilon_2$, an invariant set $\mathcal E\subset \R^n$, $\mathcal E \supset \mathcal D$,
a symmetric positive definite matrix $P$, and a matrix $K$ such that, $\forall x \in \mathcal E$,
\begin{align}\label{eq:cond_der}
&x^T \left\{Q+K^TRK+\mathrm{symm}\Bigl(P\left[A+BK\right]+\right.\nonumber \\
%\left[A+BK\right]^T P
%&\left.+P
%\begin{pmatrix}x^T (F_1+G_1 K)
% \\ \vdots \\
%x^T (F_n+G_n K)
%\end{pmatrix}
%\right.
%\nonumber \\
&+\left.
\begin{pmatrix} (F_1+G_1 K)^T x & \dots & (F_n+G_n K)^T x \end{pmatrix} P \Bigr)
\right\}x\nonumber \\
&+\epsilon_1 x^T P D D^T P x+ \epsilon_1^{-1} x^T (E_1+E_2 K)^T (E_1 + E_2 K) x\nonumber\\
&+\epsilon_2 x^T P\left[I_n\otimes\bigl(x^TD\bigr)\right]\left[I_n\otimes \bigl(D^Tx\bigr) \right] Px\nonumber\\
%&+\epsilon_2^{-1} x^T\left(\left(R_1+S_1 K\right)^T \dots \left(R_n+S_n K\right)^T \right)  \nonumber\\
%&\hspace{3.5cm}\times\begin{pmatrix}
&+\epsilon_2^{-1} x^T \bigl( (R_1+S_1 K )^T \dots (R_n+S_n K)^T \bigr)  \nonumber\\
&\left((R_1+S_1 K )^T \dots (R_n+S_n K)^T \right)^T x < 0\,.
%&\hspace{3.5cm}
%\times\begin{pmatrix}
%R_1+S_1 K
% \\ \vdots \\
%R_n+S_n K
%\end{pmatrix}x<0\,, \quad\forall x \in \mathcal E\,.
\end{align}
%
%\vspace{-0.8 cm}
Then, the state feedback controller \eqref{eq:sf} is a QGCC for the uncertain system \eqref{eq:sys} with associated  cost matrix $P$
%\begin{equation} \label{barJ}
%\bar J := \max_{x_0\in \mathcal E} x_0^TPx_0\,.
%\end{equation}
\end{lemma}
%Before proving the lemma, some considerations are in order.
%
%\vspace{-0.4 cm}

%\begin{remark} \label{rem2}
%Condition \eqref{eq:cond_der} guarantees that the derivative of the quadratic Lyapunov function~$v(x)=x^TPx$, computed  along the solutions of the closed loop system~\eqref{eq:cl_sys}, with $B_w = 0 $, is negative definite over the invariant set $\mathcal E$. Note that it is necessary to include  the admissible set $\mathcal D$ into an invariant set, since $\mathcal D$ is a problem data and, in general, {\em is not} an invariant set. From classical Lyapunov theory, the inclusion of the admissible set $\mathcal D$ into an invariant set $\mathcal E$, together with the negative definiteness of a {\em quadratic} Lyapunov function in the set $\mathcal E$, guarantees that every trajectory starting in $\mathcal D$ converges exponentially to zero, thus ensuring that the set $\mathcal D$ is included into the DES of the closed loop system (point i) in Definition \ref{defn:RGCC}).
%%\noindent
%%It is clear that, in presence of a disturbance, the state trajectories of the closed loop system will remain inside the bounded set $\mathcal E$ whenever the disturbance energy is bounded by an allowable amount; however, $\mathcal D$ no longer belongs to the DA of the closed system.
%%\hfill$\triangle$
%\end{remark}
%\vspace{-0.6 cm}
\begin{pf}
Consider the candidate Lyapunov function $v(x)=x^T P x$.
%The time derivative of $v(x)$ along the trajectories of the closed loop system is given by \eqref{eq:der_cl}.
%\footnote{In  condition \eqref{eq:der_cl} we explicitly write down the terms depending on $B_w$, although $B_w$ is assumed to be zero, because the complete expression of $\dot v$ will be useful for the proof of  the next lemma. In this way, we avoid to rewrite the expression of $\dot v$ again.}
By exploiting Lemma \ref{lemma:petersen}, it is straightforward to prove the following majoration holds
\begin{align} \label{eq:v_x_maj_2}
\dot v(x) \leq & x^T \left\{ \mathrm{symm}\left( P\left[A+BK\right] \right. \right. \nonumber \\
    & \left.\left. + \begin{pmatrix}  (F_1+G_1 K)^T x & \dots & (F_n+G_n K)^T x \end{pmatrix} P
\right) \right\} x \nonumber \\
& + \mathrm{symm}\left( x^T P B_w w \right) + \epsilon_1 x^T P D D^T P x  \nonumber\\
& + \epsilon_1^{-1} x^T (E_1+E_2 K)^T (E_1 + E_2 K) x \nonumber \\
& +\epsilon_2 x^T P \bigl(I_n\otimes x^TD \bigr)\bigl(I_n \otimes D^Tx\bigr)P x \nonumber \\
& +\epsilon_2^{-1}x^T\left(\left(R_1+S_1 K\right)^T \dots \left(R_n+S_n K\right)^T \right)\nonumber \\
& \left(\left(R_1+S_1 K\right)^T \dots \left(R_n+S_n K\right)^T \right)^T x\,.
%\begin{pmatrix}
%R_1+S_1 K
% \\ \vdots \\
%R_n+S_n K
%\end{pmatrix}x\,.
\end{align}

In view of \eqref{eq:v_x_maj_2}, since ${B_w=0}$, condition \eqref{eq:cond_der} yields
\begin{align}\label{eq:cond1}
\dot v(x)<-x^T(Q+K^TRK)x\,, \quad  \forall x \in \mathcal E\,.
\end{align}
%For $w(t)\equiv 0$, \eqref{eq:cond_der}
Condition \eqref{eq:cond1} guarantees the negative definiteness of $\dot v(x)$ over the invariant set $\mathcal E$.
Therefore, using standard Lyapunov arguments, %(see \cite{Khalil}, Ch. 4),
it is possible to conclude that the equilibrium point $x=0$ is
exponentially stable, whereas $\mathcal E$ is contained into the DES of the equilibrium point of the closed loop system.
Hence, each trajectory starting from an arbitrary $x_0 \in \mathcal E$ converges exponentially to zero; therefore, for each $x_0\in\mathcal E$, it makes sense to
integrate both sides of \eqref{eq:cond1} from $0$ to $+\infty$; we obtain
%\begin{equation}
%- x^T_0Px_0< -\int_0^\infty \left(x^T(t)Qx(t) +u^T(t)Ru(t)\right)dt<0 \,. \label{eq_int}
%\end{equation}
%
%Therefore, from \eqref{eq_int},
\begin{equation} \label{eq_ind_bound}
J_2<x^T_0Px_0\,.
\end{equation}

The proof follows from the arbitrariness of $x_0$, and
 the fact that $\mathcal D$ is included into the invariant set $\mathcal E$.\hfill$\blacksquare$
\end{pf}

Now let us consider the $\mathcal L_2$-performance control problem; the following technical lemma is necessary for the derivation of the main result.
\begin{lemma}\label{lemma:H_inf}
Consider the uncertain NQS \eqref{eq:sys}.%, and the admissible uncertainties satisfying \eqref{eq:def_NBUs}-\eqref{eq:def_F}.
Given the cost index~\eqref{eq:cost_idx_H_inf}, assume there exist some positive scalars $\epsilon_1$, $\epsilon_2$,
 a symmetric positive definite matrix $P$, and a matrix $K$ such that, %for any matrix $H$ satisfying \eqref{eq:def_F},
\begin{itemize}
\item[i)] The reachable set $\mathcal R_{CL}$  of  the closed loop system \eqref{eq:cl_sys} is a finite subset of $\R^n$ ;
\item[ii)] The following inequality holds for all $x \in \mathcal R_{CL}$, and $w\in\R^{n_w}$,
\begin{align}
& \hspace{-1.45cm} x^T \left\{C^T C + \mathrm{symm} \left( P \left[ A + BK \right] \right) \right.\nonumber\\
& \left. + \mathrm{symm}\left( \begin{pmatrix} (F_1+G_1 K)^T x & \dots & (F_n+G_n K)^T x \end{pmatrix} P \right)
\right\} x  \nonumber\\
& +\epsilon_1 x^T P D D^T P x+ \epsilon_1^{-1} x^T (E_1+E_2 K)^T (E_1 + E_2 K) x \nonumber\\
& +\epsilon_2 x^T P \bigl(I_n\otimes x^TD\bigr) \bigl(I_n \otimes D^Tx\bigr)P x \nonumber\\
& + \epsilon_2^{-1} x^T \left(\left(R_1+S_1 K\right)^T \dots \left(R_n+S_n K\right)^T \right) \nonumber\\
& \left(\left(R_1+S_1 K\right)^T \dots \left(R_n+S_n K\right)^T \right)^T x  \nonumber\\
& + \mathrm{symm}\left(x^T P B_w w\right) - w^T w<0\,. \label{eq:cond_der_H_inf}
\end{align}
\end{itemize}
Then, the state feedback controller \eqref{eq:sf} is a G$\mathcal L_2$PC for the NQS \eqref{eq:sys}.
\hfill$\blacksquare$
\end{lemma}
\begin{pf}
Let us consider the candidate Lyapunov function $v(x)=x^T P x$ %The time derivative of $v(x)$ along the trajectories of the closed loop system is given by \eqref{eq:der_cl}.
and the $\mathcal H_\infty$ performance index~\eqref{eq:cost_idx_H_inf}. $J_\infty$ satisfies
\begin{align} \label{D1}
J_\infty &= \int_0^{\infty} \left(z(t)^T z(t) - w(t)^T w(t) + \dot v(x(t))\right) dt \nonumber\\
&- \int_0^{\infty} \dot v(x(t)) dt \,.
\end{align}

Since $\mathcal R_{CL}$ is a finite subset of $\R^n$, we have that $x(\cdot)$ is bounded at infinity; moreover $x(0)=0$, therefore
\begin{equation}\label{eq:liminf}
\int_0^{\infty} \dot v(x(t)) dt \geq \liminf_{t\to\infty} v(x(t)) \geq 0\,.\\
\end{equation}

From \eqref{D1} and \eqref{eq:liminf} we obtain
\begin{equation}\label{eq:widehat_ineq}
J_\infty \leq\int_0^{\infty} (z^T(t) z(t) - w^T (t)w (t)+ \dot v(x(t)) ) dt \,.
\end{equation}

From \eqref{eq:widehat_ineq}, a sufficient condition for negative definiteness of $J_\infty$ is %negative if the following inequality holds
\begin{align}\label{eq:widehat_ineq_2}
\dot v(x(t)) \leq -z^T(t)z(t) + w(t)^T w(t) \,,\quad\forall t\in[0,+\infty)\,.
\end{align}

In view of \eqref{eq:v_x_maj_2}, condition~\eqref{eq:cond_der_H_inf} implies the satisfaction of~\eqref{eq:widehat_ineq_2} for all $t\geq 0$; therefore condition {\bf ii)} in Definition~\ref{defn:H_inf} is satisfied. Moreover, if $w=0$, condition~\eqref{eq:widehat_ineq_2} guarantees negative definiteness of~$\dot v$ over~$\mathcal R_{CL}$, which in turn implies the satisfaction of condition~i) of Definition~\ref{defn:H_inf}.
\hfill$\blacksquare$
%After integration of both sides of \eqref{eq:widehat_ineq_3} from $0$ to $\infty,$
%\begin{align}\label{eq:widehat_ineq_4}
%v(x(\infty))-v(x(0)) & \leq -\int_0^{\infty} y^T(\sigma) y(\sigma) d\sigma\nonumber \\ & + \mu^2 \int_0^{\infty} w^T(\sigma) w(\sigma) d\sigma\,.
%\end{align}
%By the positive definiteness of the Lyapunov function and taking into account again the hypothesis of zero initial condition $v(0)=0$,
%\eqref{eq:widehat_ineq_4} is equivalent to
%\[
%v(x(0))=0\,,
%\]
%since $v(\cdot)$ is positive definite
%\[
%\int_0^\infty y^T(\sigma)y(\sigma) d\sigma \leq \mu^2 \int_0^\infty w^T(\sigma) w(\sigma) d\sigma\,;
%\]
%from the last inequality the proof follows.
\end{pf}

%%
%The following remark clarifies the relationship between the admissible set $\mathcal D$ and the reachable set $\mathcal R_{CL}$.
%%
%\begin{remark} \label{unsatisfied}
%As said in the proof of Lemma \ref{lemma:H_inf}, we have that $\dot{v}$ is negative definite over the reachable set $\mathcal R_{CL}$; therefore, by using the same machinery of the proof of Lemma \ref{lemma:bound}, it is simple to recognize that $\mathcal R_{CL}$ belongs to the DES of the closed loop system. Since the set $\mathcal R_{CL}$ is not {\em a priori} assigned - it is only required that it is a finite subset of $\R^n$, to guarantee boundedness of the trajectories at infinity -
%it might not be a satisfactory estimate of the DES from the designer point of view. In other words, if it is desired that a given region $\mathcal D$ belongs to the DES of the closed loop system, and $\mathcal D$ is not a subset of $\mathcal R_{CL}$, the negative definiteness of $\dot v$ over $\mathcal D$ must be guaranteed by a further LMI constraint, following the same guidelines of Lemma \ref{lemma:bound}.
%\end{remark}

%%%%%%%%%%%%%%%%%%%%%%%%%%%%%%%%%%%
%\subsection{Polytopic admissible sets}
%%%%%%%%%%%%%%%%%%%%%%%%%%%%%%%%%%%%
In the main results, given in the next section, we will assume that the admissible set $\mathcal D$ is a polytope. Therefore, let us recall that a polytope $\mathcal P\subset \mathbb{R}^n$ can be described
as follows
\begin{subequations} \label{polytope}
\begin{align}
\mathcal P&= \co\left\{x_{(1)},x_{(2)},\dots,x_{(r)}\right\} \label{descr_R_co}\\
& = \left\{x\in\mathbb{R}^n\,:\, a_k^T x \leq
1\,,\,k=1,2,\dots,q\right\}\,,\label{descr_R_bounds}
\end{align}
\end{subequations}
where $p$ and $r$ are suitable integers, $x_{(i)}$ denotes the
$i$-th vertex of the polytope $\mathcal P$, $a_k \in \mathbb R^n$
and $\co\{\cdot\}$ denotes the operation of taking the convex hull
of the argument.

%%%%%%%%%%%%%%%%%%%%%%%%%%%%%%%%%%%%%%%%%%%%%%%%%%%%%%%%%%%%%%%%%%%%%%%%%%%%%%%%%%%%%%%%%%%%%%%%%%%%%%%%%%%%%%%%%%%%%%%%%%
\section{Main Results}\label{sec:main}
%%%%%%%%%%%%%%%%%%%%%%%%%%%%%%%%%%%%%%%%%%%%%%%%%%%%%%%%%%%%%%%%%%%%%%%%%%%%%%%%%%%%%%%%%%%%%%%%%%%%%%%%%%%%%%%%%%%%%%%%%%
The next theorems state some sufficient conditions for the existence of QGCCs and G$\mathcal L_2$PCs for uncertain NQSs with external disturbance and constraints on the control input.
For the further developments, it is assumed that  the admissible set  has a polytopic structure; therefore we let $\mathcal D=\mathcal P$.

%%%%%%%%%%%%%%%%%%%%%%%%%%%%%%%%%%%%%%%%%%%%%%%%%%%%%%%%%%%%%%%%%
%MAIN THEOREM
%%%%%%%%%%%%%%%%%%%%%%%%%%%%%%%%%%%%%%%%%%%%%%%%%%%%%%%%%%%%%%%%%
\subsection{Design of QGCCs} \label{sec:qgcc}
%%MAIN THEOREM
%%%%%%%%%%%%%%%%%%%%%%%%%%%%%%%%%%%%%%%%%%%%%%%%%%%%%%%%%%%%%%%%%
\begin{thrm}\label{th:th_main}
Given the uncertain system \eqref{eq:sys},  an admissible polytopic set $\mathcal P$ in the form \eqref{polytope}, some positive scalars $u_{i,max}$, \mbox{$i=1,\dots,m$}, the cost index \eqref{eq:cost_idx}, if there exist some positive scalars $\epsilon_1,\epsilon_2$, a scalar $\gamma$, a matrix $Y$ and symmetric positive definite matrices $X$ such that
\begin{subequations}
\begin{align}
0<\gamma &< 1 \label{th_main:a}\\
\begin{pmatrix} 1  & \gamma a_k^T X \\ X a_k \gamma & X \end{pmatrix} &\geq 0\,, \quad k=1,2,\dots,q \label{th_main:b}\\
\begin{pmatrix} 1 & x_{(i)}^T \\ x_{(i)} & X \end{pmatrix} &\geq 0\,, \quad i=1,2, \dots, r  \label{th_main:c}\\
\begin{pmatrix} U_{\mathrm{max}}^2 & Y \\ Y^T & X \end{pmatrix} &\geq 0\,,\label{th_main:d}\\
\begin{pmatrix}
              L_{(i)} & \gamma X & \gamma Y^T & \gamma W^T & \gamma M^T & \Gamma^T_{(i)}\\
              \gamma X & -\gamma Q^{-1} & 0 & 0 &0 & 0\\
               \gamma Y & 0 & -\gamma R^{-1}  & 0 & 0 & 0 \\
              \gamma W & 0 & 0 & -\gamma \epsilon_1 I & 0 & 0\\
              \gamma M & 0 & 0 & 0 & -\gamma \epsilon_2 I & 0\\
              \Gamma_{(i)} & 0 & 0 & 0 & 0 & -\gamma \epsilon_2 I
              \end{pmatrix}&< 0 \,,\quad  i=1,2, \dots,r\,.  \label{th_main:f}
\end{align}
\end{subequations}
where $a_k$, $k=1,2,\dots,q$, $x_{(i)}$, $i=1,2, \dots, r$ and $U_{\mathrm{max}}=\mathrm{diag}(u_{1,\mathrm{max}},\dots,u_{m,\mathrm{max}})$, define the polytope $\mathcal P$ according to \eqref{polytope}, and
\begin{align*}
L_{(i)} := & \gamma \, \mathrm{symm}(AX+BY) + \gamma \epsilon_1 DD^T\\
& \hspace{-1.15cm} + \mathrm{symm}\begin{pmatrix} (F_1 X + G_1 Y)^T  x_{(i)} &
\dots & (F_n X + G_n Y)^T  x_{(i)} \end{pmatrix}\,,\\
W:=&E_1X+E_2Y\,,\quad M:=\begin{pmatrix} R_1 X + S_1 Y \\
\vdots \\ R_n X + S_n Y  \end{pmatrix}\,,\nonumber\\
\Gamma_{(i)}:=& \epsilon_2 \bigl(I_n \otimes D^Tx_{(i)} \bigr)\,,
\end{align*}
then $u(t)=YX^{-1}x(t)$ is  a QGCC for system \eqref{eq:sys} with associated cost matrix $X^{-1}$, and satisfying the
input constraints \eqref{eq:cstr_u}.
\end{thrm}
\begin{pf}
Given the scalar $\gamma$ satisfying the hypothesis of the theorem, let $\rho=1/\gamma >1$ and define $\rho\, \mathcal P$ as the polytope
obtained by multiplying by $\rho$ the coordinates of the vertices of $\mathcal P$.
After multiplying \eqref{th_main:f}  by $\rho$, all of its elements become affine matrix functions of the variable $x$. Therefore, it is possible to invoke the result in ~\cite{Amato}, Ch.~3, which guarantees that an affine function is negative definite on the polytope $\rho \mathcal P$ if and only if the property holds at the vertices of the polytope. Thus \eqref{th_main:f} can be equivalently rewritten as
\begin{align}\label{eq:cond_der_th4}
&\begin{pmatrix}
              \Xi(\rho x_{(i)}) & X & Y^T & W^T & M^T & \Pi^T(\rho x_{(i)})\\
               X & -Q^{-1} & 0 & 0 &0 & 0\\
               Y & 0 & -R^{-1}  & 0 & 0 & 0\\
              W & 0 & 0 & -\epsilon_1 I & 0 & 0\\
              M & 0 & 0 & 0 & -\epsilon_2 I & 0\\
              \Pi(\rho x_{(i)}) & 0 & 0 & 0 & 0 & -\epsilon_2 I
              \end{pmatrix}<0\,,
\end{align}
where
\begin{align*}
\Xi(x) &:= \, \mathrm{symm}(AX+BY) \nonumber\\
&\quad+ \epsilon_1 DD^T + \mathrm{symm}\begin{pmatrix} (F_1 X + G_1 Y)^T  x &
\dots & (F_n X + G_n Y)^T  x \end{pmatrix}\,,\\
\Pi(x) &:= \epsilon_2 \left(I_n \otimes D^T x\right)
\end{align*}
By noting that the matrix functions $\Xi(\cdot)$ and $\Pi(\cdot)$ depend affinely on their arguments, it is possible to invoke the result in \cite{Amato}, Ch.3, which guarantees that an affine function is negative definite on the polytope $\rho \mathcal P$ if and only if the property holds at the vertices of the polytope. Therefore, using also the properties of the Schur complements (see \cite{Boyd}, p.7), condition
\eqref{eq:cond_der_th4} is equivalent to
\begin{align}\label{eq:cond_der_th3}
X Q X &+ Y^T R Y + \mathrm{symm}\left(AX+BY\right) \nonumber \\
& + \mathrm{symm}\begin{pmatrix} (F_1X+G_1Y)^T x & \dots & (F_nX+G_nY)^T x \end{pmatrix}\nonumber \\
& +\epsilon_1 DD^T+\epsilon_1^{-1}(E_1X+E_2Y)^T (E_1X+E_2Y) \nonumber \\
& +\epsilon_2^{-1} \begin{pmatrix} (R_1X+ S_1Y)^T  & \dots & (R_nX+S_nY)^T  \end{pmatrix}
             \begin{pmatrix} (R_1X+S_1Y) \\
                                       \vdots \\
                                      (R_nX+ S_nY)
                                         \end{pmatrix} \nonumber \\
& +\epsilon_2 \bigl(I_n \otimes D^T x \bigr)^T\bigl(I_n \otimes D^T x \bigr) <0\,,\quad \forall x \in \rho \mathcal P\,.
\end{align}
Pre- and post- multiplying the left-hand side of \eqref{eq:cond_der_th3} by $ X^{-1}=:P$,  and letting $K:=YP$, \eqref{eq:cond_der_th3} can be rewritten as \eqref{eq:cond_der}.
The completion of the proof can be achieved through the following steps.
\begin{itemize}
\item[\textbf{i)}] Letting $X= P^{-1}$ in \eqref{th_main:c}, from the result in~\cite{Boyd}, p. 69, condition \eqref{th_main:c} ensures the inclusion of
the polytope $\mathcal P$ into the ellipsoid
\begin{equation}   \label{E}
\tilde{\mathcal E} =\left\{ x\in \R^n \,,\, x^T P x \leq  1 \right\}\,.
\end{equation}
\item[\textbf{ii)}] Using again the Schur complements and recalling that $X=  P^{-1}$ and $\gamma=1/\rho$,
\eqref{th_main:b} is equivalent to
\begin{equation}
\frac{a_k^T}{\rho}  P^{-1}\frac{a_k}{\rho} \leq 1\,, \quad
k=1,2,\dots,q\,,
\end{equation}
which implies $\rho \mathcal P \supset \tilde{\mathcal E}\supset \mathcal P$ (see \cite{Boyd}, p. 70).
%Such an inclusion guarantees that $\dot v(x)$ is negative definite on the invariant set $\mathcal E$. Therefore, the level curves
%of $v(x)$, which are bounded by $\mathcal E$, are decreasing along the system trajectories.

Therefore we can conclude that there exists an invariant set $\tilde {\mathcal E}$, containing the admissible set $\mathcal P$, such that condition \eqref{eq:cond_der} is satisfied  on $\tilde {\mathcal E}$. Hence, the application of Lemma~\ref{lemma:bound} allows to conclude that $u(t) = YX^{-1}x(t)$ is a QGCC for system \eqref{eq:sys} with associated cost matrix $P$. Moreover:
\item[\textbf{iii)}] Recalling that $X=P^{-1}$ and $Y=KX$, \eqref{th_main:d} is equivalent to
\begin{equation}\label{eq:Z}
\begin{pmatrix} U_\mathrm{max}^2 & K  P^{-1} \\  P^{-1} K^T &  P^{-1} \end{pmatrix} \geq 0\,.\
\end{equation}
The Schur complements of \eqref{eq:Z} yield $K P^{-1} K^T \leq U_\mathrm{max}^2\,$.
Therefore, denoting the $i$-th row of the matrix $K$ by $k_i$, we have that, for all $x\in \mathcal P$,
\begin{align}\label{eq:cond_norm_u}
|u_i|^2=&|k_i x|^2=|k_i P^{-1/2} P^{1/2} x|^2\nonumber\\
&\leq \|k_i P^{-1/2}\|^2 \|P^{1/2} x\|^2\nonumber\\
&=k_i P^{-1} k_i^T x^T(t) P x\nonumber\\
&\leq k_i P^{-1} k_i^T \leq u_{i,\mathrm{max}}^2\,.
\end{align}
Inequality \eqref{eq:cond_norm_u} allows to conclude that the control law \eqref{eq:sf},
with $K=YX^{-1}$, also satisfies the input constraints \eqref{eq:cstr_u}; this concludes the proof.\hfill$\blacksquare$
\end{itemize}
\end{pf}

Note that a minimization of the guaranteed cost can be achieved by minimizing the volume of the set~\eqref{E}, through
its approximated measure provided by~$\trace(P^{-1})$.
Since, for a given $\gamma \in (0,1)$, the conditions of Theorem~\ref{th:th_main} are a set of Linear Matrix Inequalities (LMIs) \cite{Boyd}
in the variables $\epsilon_1$, $\epsilon_2$, $X$, $Y$, which can be solved via available software \cite{Optim_tool}, we propose the following convex optimization problem, for a fixed $\gamma$. A one parameter search in order to optimize $\mathcal \gamma$ over the interval $(0,1)$ is necessary.
%In particular, given $\gamma \in (0,1)$, the design problem for the optimal guaranteed cost controller can be
%formulated through the following convex optimization problem.
%\begin{prob}\label{prob_min_J}
%\begin{align}
%%&\min_{\bar J, \epsilon_1, \epsilon_2, X, Y, Z, u_{i,max,i=1,\dots,m}} \quad \bar J\nonumber \\
%%&\min_{\bar J, \epsilon_1, \epsilon_2, X, Y, Z, u_{i,max,i=1,\dots,m}} \quad \bar J\nonumber \\
%&\min_{J,\epsilon_1, \epsilon_2, X, Y, Z} \quad J\nonumber \\
%&s. t. \:
%\eqref{cond2_J}\,,\eqref{th_main:b}\,, \eqref{th_main:c}\,, \eqref{th_main:d}\,, \eqref{th_main:e}\,,\eqref{th_main:f}\nonumber.
%\end{align}
%\end{prob}
\begin{problem}\label{prob_min_J}
\begin{align}
%&\min_{\bar J, \epsilon_1, \epsilon_2, X, Y, Z, u_{i,max,i=1,\dots,m}} \quad \bar J\nonumber \\
%&\min_{\bar J, \epsilon_1, \epsilon_2, X, Y, Z, u_{i,max,i=1,\dots,m}} \quad \bar J\nonumber \\
&\min_{\epsilon_1, \epsilon_2, X, Y} \trace(X) \\
&s. t. \:
\eqref{th_main:b}\,, \eqref{th_main:c}\,, \eqref{th_main:d}\,, \eqref{th_main:f}\nonumber.
\end{align}
\end{problem}
%If Problem \ref{prob_min_J} has an optimal solution, then $u(t)=YX^{-1}x(t)$ is a QGCC with associated cost matrix $P=X^{-1}<JI$, for the NQS \eqref{eq:sys}, satisfying the control input constraints \eqref{eq:cstr_u}.
%
If Problem \ref{prob_min_J} has an optimal solution, then $u(t)=YX^{-1}x(t)$ is a QGCC with associated cost matrix $P=X^{-1}$, for the NQS \eqref{eq:sys}, satisfying the control input constraints \eqref{eq:cstr_u}.

%%%%%%%%%%%%%%%%%%%%%%%%%%%%%%%%%%%%%%%%%%%%%%%%%%%%%%%
\subsection{Design of G$\mathcal L_2$PCs } \label{sec:gL2pc}
%%%%%%%%%%%%%%%%%%%%%%%%%%%%%%%%%%%%%%%%%%%%%%%%%%%%%%%
%Robust Guaranteed Cost Control of Linear Systems with Norm Bounded Uncertainties and External Disturbance
A solution to the G$\mathcal L_2$PC design problem, which allows to address both robustness
constraints and disturbance attenuation requirements, is proposed through the following theorem.
\begin{thrm}\label{th:th_H_inf}
Given the uncertain system \eqref{eq:sys}, the polytope $\mathcal P_\infty \subset \mathbb R^n$, defined according to \eqref{polytope}, %a cost function \eqref{eq:cost_idx_H_inf}
and some positive scalars $u_{i,max}$, $i,\dots,m$,  if there exist  positive scalars $\alpha$, $\epsilon_1$, $\epsilon_2$,
 a matrix $Y$, and a symmetric positive definite matrix $X$ such that
{\small
\begin{subequations}
\begin{align}
%0<\gamma &< 1 \label{th_H_inf:a}\\
\begin{pmatrix} 1  & a_k^T X \\ X a_k & X \end{pmatrix} &\geq 0\,, \quad k=1,2,\dots,q, \label{th_H_inf:b}\\
%\begin{pmatrix} 1 & x_{(i)}^T \\ x_{(i)} & X \end{pmatrix} &\geq 0\,, \quad i=1,2, \dots, r  \label{th_H_inf:c}\\
\begin{pmatrix} U_{\mathrm{max}}^2 & Y \\ Y^T & X \end{pmatrix} &\geq 0\,,\label{th_H_inf:d}\\
%(Z_{ii})&\leq u_{i,max}^2\,, \quad i=1,2, \dots, m\,, \label{th_H_inf:e}\\
\left(
\begin{matrix}
              L_{(i)} & W^T & X C^T & M^T & \Gamma^T_{(i)} & B_w \\
              W & -\epsilon_1 I & 0 & 0 & 0 & 0\\
              C X & 0 &- I & 0 & 0 & 0\\
              M & 0 & 0 & -\epsilon_2 I & 0 & 0\\
              \Gamma_{(i)} & 0 & 0 & 0 & -\epsilon_2 I & 0\\
              B^T_w & 0 & 0 & 0 & 0 & -I
              \end{matrix}
\right) &< 0 \,, \label{th_H_inf:f} \qquad i=1,2, \dots,r\,.
\end{align}
\end{subequations}
}%end small
where $a_k$, $k=1,2,\dots,q$, $x_{(i)}$, $i=1,2, \dots, r$ and
$U_{\mathrm{max}}=\mathrm{diag}(u_{1,\mathrm{max}},\dots,u_{m,\mathrm{max}})$,
define the  polytope $\mathcal{P_{\infty}}$ according to \eqref{polytope}, and
\begin{align*}
L_{(i)} &:= \mathrm{symm}\left(AX+BY\right) + \epsilon_1 DD^T\\
& + \mathrm{symm}\begin{pmatrix} (F_1 X + G_1 Y)^T  x_{(i)} &
\dots & (F_n X + G_n Y)^T  x_{(i)} \end{pmatrix}\,,\\
W &:= E_1X+E_2Y\,,\quad M:=\begin{pmatrix} R_1 X + S_1 Y \\
\vdots \\ R_n X + S_n Y  \end{pmatrix}\,,\nonumber\\
\Gamma_{(i)} &:= \epsilon_2 \bigl(I_n \otimes D^T x_{(i)} \bigr)\,,
%&\hspace{5.5cm}i=1,2, \dots,r\,.
\end{align*}
then $u(t)=YX^{-1}x(t)$ is a G$\mathcal L_2$PC, for system \eqref{eq:sys}, satisfying the input constraints \eqref{eq:cstr_u}.
\end{thrm}
\begin{pf}
Consider the quadratic Lyapunov function candidate for system~\eqref{eq:cl_sys} as
$v(x)=x^T P x$. Letting $P= X^{-1}$, and $K=YX^{-1}$, the proof proceeds through similar arguments of the proof of Theorem~\ref{th:th_main}.
Using both the properties of norm-bounded uncertainties (see Lemma~\ref{lemma:petersen}) and Schur complements,
and after exploiting the affine structure of the resulting matrix function,~\eqref{th_H_inf:f} is readily seen to imply~\eqref{eq:cond_der_H_inf}; therefore condition~ii) in Lemma~\ref{lemma:H_inf} is satisfied over the set~${\mathcal P_\infty}$.

By \eqref{th_H_inf:b}, and through the LMI conditions in \cite{Boyd}, p.70, it readily follows that the ellipsoid $\tilde {\mathcal E}$ defined in \eqref{E} is such that
\begin{equation}\label{E_tilde}
{\tilde{\mathcal E}} \subset \mathcal P_\infty \,.
\end{equation}
%\end{equation}
%is contained into $\mathcal P_\infty$.

Now we shall prove that the ellipsoid~$\tilde {\mathcal E}$ contains the reachable set $\mathcal R_{CL}$ of the closed loop system. Indeed, since   condition \eqref{th_H_inf:f} implies~\eqref{eq:cond_der_H_inf}, we have that, for all $x\in \tilde {\mathcal E}$ and $w$,
\begin{align}
\dot v(x) & \leq -z^T z + w^T w \nonumber \\
                 & \leq  w^T w \,. \label{old_ineq}
\end{align}

Integration of both sides of \eqref{old_ineq} between $0$ and $t>0$, yields
\begin{equation}
x^T(t)Px(t)  \leq \int_0^t w^T(\sigma)w(\sigma) d\sigma \le 1\,;
\end{equation}
therefore, we can conclude that
\begin{equation} \label{inclusions}
\mathcal R_{CL} \subset  \tilde {\mathcal E}  \subset \mathcal P_\infty\,.
\end{equation}

From the first inclusion in \eqref{inclusions} it follows condition~i) in Lemma~\ref{lemma:H_inf}, while the second inclusion guarantees the satisfaction of condition ii). Finally, inequality \eqref{th_H_inf:d}, as in Theorem \eqref{th:th_main}, ensures condition \eqref{eq:cstr_u}.
This completes the proof.\hfill$\blacksquare$
\end{pf}
%\begin{remark}
%The set $\mathcal P_\infty$ contains the sets $\tilde {\mathcal E}\supset\mathcal R_{CL}$. According to Remark~\ref{unsatisfied}, both sets are estimates, possibly not satisfactory, of the DES of the closed loop system. As explained in Remark \ref{unsatisfied}, according to the designer requirements, it could be necessary a further LMI constraint to guarantee a larger estimate of the DES of the closed loop system.
%%\hfill$\triangle$
%\end{remark}

% In the following development, a LMI optimization problem is formulated in order to minimize the gain $\mu$.
%
%%Therefore, the following LMI problem can be formulated.
% %\hfill$\triangle$
% %\end{remark}
%
% \begin{prob}\label{prob_min_H_inf}
%\begin{align}
%& \min_{\mu, \epsilon_1, \epsilon_2, X, Y, Z} \quad \mu\nonumber \\
%&s. t.\:
%\eqref{th_H_inf:b}\,, \eqref{th_H_inf:d}\,, \eqref{th_H_inf:e}\,, \eqref{th_H_inf:f}\,.
% \end{align}
%\end{prob}
%If Problem \ref{prob_min_H_inf} has an optimal solution, $u(t)=YX^{-1}x(t)$ is a G$\mathcal L_2$PC for system \eqref{eq:sys}, with guaranteed gain $\mu$, satisfying the control input constraints.
%
%\textbf{Alternatively, given $\mu$, it is possible to maximize the bound of the admissible disturbance energy for an assigned $\mathcal P_\infty$. This amounts to find the maximum volume ellipsoid \eqref{E} contained into the set $\mathcal P_\infty$.}
%
%Therefore, the following LMI optimization problem is proposed.

Robust control performance can be achieved by minimizing the reachable set bounding
all the state trajectories perturbed by the disturbance. To this end,
the next problem minimizes the volume of the ellipsoid $\tilde {\mathcal E}$ such that $\mathcal R_{CL} \subset  \tilde {\mathcal E}  \subset \mathcal P_\infty$.
 \begin{problem}\label{prob_max_reachable_set}
\begin{align}
& \max_{\epsilon_1, \epsilon_2, X, Y} \quad \trace(X)\nonumber \\
&s. t.\:
\eqref{th_H_inf:b}\,, \eqref{th_H_inf:d}\,, \eqref{th_H_inf:f}\,.
 \end{align}
\end{problem}
If Problem \ref{prob_max_reachable_set} has an optimal solution, $u(t)=YX^{-1}x(t)$ is a G$\mathcal L_2$PC for system \eqref{eq:sys}, satisfying the control input constraints.

%\begin{remark}
%Problem \ref{prob_min_H_inf} provides, through the ``biggest'' level curve of the Lyapunov function contained into $\mathcal P_\infty$, an optimal estimate
%of the bound $\alpha$ and, therefore, of the energy bound $w_{max}$. (see also Remark \ref{rem:energy_bound}).
%\hfill$\triangle$
%\end{remark}

%=======================================================================================================================

%%%%%%%%%%%%%%%%%%%%%%%%%%%%
\section{Conclusions} \label{sec:concl}
%%%%%%%%%%%%%%%%%%%%%%%%%%%%
The problem of robust and optimal control for the class of NQSs subject to norm-bounded parametric uncertainties and disturbance inputs has
been investigated. Some constraints, both on control inputs and disturbance attenuation, have been also taken into account into the proposed control design methodologies which are conceived as contributions to a unified theory of constrained and optimal control for uncertain NQSs.\\
The guaranteed cost control and the $\mathcal L_2$-gain disturbance rejection problems have been addressed. A common feature of both the devised techniques is that the Lyapunov stability of the equilibrium is guaranteed and, moreover, the optimization conditions yields regions included into the DES of the equilibrium.
The proposed design methods are both effectively applicable, since they are based on the solution of a LMI optimization problem, which can be easily computed by means of off-the-shelf software packages.

%The effectiveness of the devised results has been illustrated through a practical case-study concerning both the motion regulation and interaction control of a robotic arm. The robot control system, including PAMs and pneumatic actuation valves, has been described as an uncertain NQS.
%Simulation results have shown the effectiveness of the contributed methodologies to fulfill the control design specifications
%in terms of quadratic regulator performance, control effort and disturbance rejection.
%
%%%%%%%%%%%%%%%%%%%%%%%%%%%%%
%\section*{Acknowledgments} \label{sec:Ackn}
%%%%%%%%%%%%%%%%%%%%%%%%%%%%%
%%
%The present work is supported by the European Commission, European Social Fund and “Regione Calabria” - Calabrian Regional Authority, Department 3 Sector 1.
%%
\bibliographystyle{IEEEtran}
\bibliography{biblio_GCC}

\end{document}